\documentclass{ws-ijmpcs}

\begin{document}

\markboth{Ju-Jun Xie, En Wang, Bo-Chao Liu and J. Nieves} {The role
of $N^*(2120)$ in the $\Lambda(1520)$ photon and hadronic
productions}

%
\catchline{}{}{}{}{}
%

\title{The role of $N^*(2120)$ nucleon resonance in $K\Lambda(1520)$ photon and hadronic productions}

\author{Ju-Jun Xie$^{1,4}$, En Wang$^2$, Bo-Chao Liu$^{3,4}$, and J. Nieves$^2$}

\address{$^1$Institute of Modern Physics, Chinese Academy of Sciences, Lanzhou 730000, China \\
$^2$Instituto de F\'\i sica Corpuscular (IFIC), Centro Mixto
CSIC-Universidad de Valencia, Institutos de Investigaci\'on de
Paterna, Aptd. 22085, E-46071 Valencia, Spain \\
$^3$Department of Applied Physics, Xi'an Jiaotong University, Xi'an,
Shanxi 710049, China \\
$^4$State Key Laboratory of Theoretical Physics, Institute of
Theoretical Physics, Chinese Academy of Sciences, Beijing 100190,
China}

 \maketitle

\begin{history}
\end{history}

\begin{abstract}

The associate $K\Lambda(1520)$ photon and hadronic production in the
$\gamma p \to K^+\Lambda(1520)$, $p p \to p K^+ \Lambda(1520)$ and
$\pi^- p \to K^0 \Lambda(1520)$ reactions are investigated within
the effective Lagrangian approach and the isobar model. We are
interested in the contribution from the $N^*(2120)$ (previously
called $N^*(2080)$) resonance, which has a significant coupling to
the $K\Lambda(1520)$ channel. The theoretical results show that the
current experimental data for the $\gamma p \to K^+\Lambda(1520)$
reaction favor the existence of the $N^*(2120)$ resonance, and that
these measurements can be used to further constrain its properties.
We present results, including the $N^*(2120)$ contribution, for
total cross sections of the $\gamma p \to K^+\Lambda(1520)$, $\pi^-
p \to K^0 \Lambda(1520)$, and $p p \to p K^+ \Lambda(1520)$
reactions. For this latter one, we also calculate invariant mass and
Dalitz plot distributions.


\end{abstract}

\ccode{PACS numbers:}

\section{Introduction}

The investigation of the baryon spectrum and the baryon couplings
from experimental data are two of the most important issues in
hadronic physics and they are attracting much attention. Both on the
experimental and theoretical sides, the nucleon excited states below
$2.0$ GeV have been extensively studied~\cite{pdg2012}. However, the
current information for the properties of states around or above
$2.0$ GeV is scarce~\cite{pdg2012}. On the other hand in this region
of energies, many theoretical approaches (constituent
quark~\cite{capstick2000} and chiral
unitary~\cite{Gamermann:2011mq,Sarkar:2009kx,Oset:2009vf,PavonValderrama:2011gp,Sun:2011fr})
predict predicted {\it missing $N^*$ states}, which have not been so
far observed. Hence, the study of the possible role played by the
$2.0$ GeV region nucleon resonances in the available accurate data
is timely and could shed light into the complicated dynamics that
governs the high excited nucleon spectrum.

The associate $K \Lambda(1520)$ photon and hadronic production
reactions might be adequate to study the $N^*$ resonances around
$2.0$ GeV, as long as they have significant couplings to the
$K\Lambda(1520)$ pair. This is because the $K\Lambda(1520)$ is a
pure isospin $1/2$ channel and the threshold is about $2.0$ GeV.
Besides, these reactions require the creation of an $\bar{s}s$ quark
pair. Thus, a thorough and dedicated study of the strangeness
production mechanism in these reactions has the potential to gain a
deeper understanding of the interaction among strange hadrons and
also on the nature of the nucleon resonances.

Recently, there have been several measurements for the $\gamma p \to
K^+\Lambda(1520)$
reaction~\cite{leps1,leps2,Wieland:2011zz,Moriya:2013hwg}. These
data suggest that there is a sizeable contribution of total and
differential cross sections from the nucleon resonances with masses
around $2.1$ GeV. On the theoretical side, in addition to the
contributions from $K$ and $K^*$ exchange in the $t-$channel and the
contact term, the contributions from the nucleon states, including
the $N^*(2120)$, in the
$s-$channel~\cite{nam2,xiejuan,hejun,nam3,xieenjuan} and the
$\Lambda(1115)$ pole in the $u-$channel have been
studied~\cite{xieenjuan}. The theoretical results show that when the
contributions from the $N^*(2120)$ resonance and the $\Lambda(1115)$
are taken into account, the current experimental
data~\cite{leps1,leps2,Moriya:2013hwg} can be well described. Thus,
it is becoming clear that the current experimental data for
$\Lambda(1520)$ photoproduction favor the existence of the
$N^*(2120)$ resonance, and that these measurements can be used to
further constrain its properties. On the other hand, based on the
results of $\gamma p \to K^+\Lambda(1520)$ reaction, the
$K\Lambda(1520)$ production in the $pp \to p K^+\Lambda(1520)$ and
$\pi^- p \to K^0 \Lambda(1520)$ hadronic processes are also studied,
paying an special attention to the contributions from the
$N^*(2120)$ resonance~\cite{xieliu}. In the present work, we will
review the main results from these theoretical studies.

\section{Study on $\gamma p \to K^+ \Lambda(1520)$ reaction}

For the $\gamma p \to K^+ \Lambda(1520)$ reaction, the differential
cross section, in the center of mass frame (c.m.), and for a
unpolarized photon beam reads,
\begin{eqnarray}
\frac{d\sigma}{d(\cos\theta_{\rm c.m.})}  &=& \frac{|\vec{k}_1^{\rm
\,\, c.m.}||\vec{p}_1^{\rm \,\,c.m.}|}{4\pi}\frac{M_N
M_{\Lambda^*}}{(W^2 - M_N^2)^2} \sum_{s_p, s_\Lambda^*,\lambda}
|T|^2, \nonumber
\end{eqnarray}
with $W$ the invariant mass of the $\gamma p$ pair. Besides,
$\vec{k}_1^{\rm \,\, c.m.}$ and $\vec{p}_1^{\rm \,\, c.m.}$ are the
photon and $K^+$ meson c.m. three-momenta, and $\theta_{\rm c.m.}$
is the  $K^+$ polar scattering angle. The invariant scattering
amplitudes are defined as
\begin{equation}
-iT_i=\bar u_\mu(p_2,s_{\Lambda^*}) A_i^{\mu \nu} u(k_2,s_p)
\epsilon_\nu(k_1,\lambda), \nonumber
\end{equation}
where $u_\mu$ and $u$ are dimensionless Rarita-Schwinger and Dirac
spinors for final $\Lambda(1520)$ and the initial proton,
respectively, while $\epsilon_\nu(k_1,\lambda)$ is the photon
polarization vector. Besides, $s_p$ and $s_{\Lambda^*}$ are the
baryon polarization variables. The sub-index $i$ stands for the
contact, $t-$channel $K^-$ exchange, contact, $s-$channel nucleon
pole and $N^*$ resonance terms (depicted in Fig.~1 of
Ref.~\cite{xiejuan}) and the $u-$channel $\Lambda(1115)$
contribution (see Fig.~2 of Ref.~\cite{xieenjuan}).

\begin{figure}
\begin{center}
\includegraphics[scale=0.23]{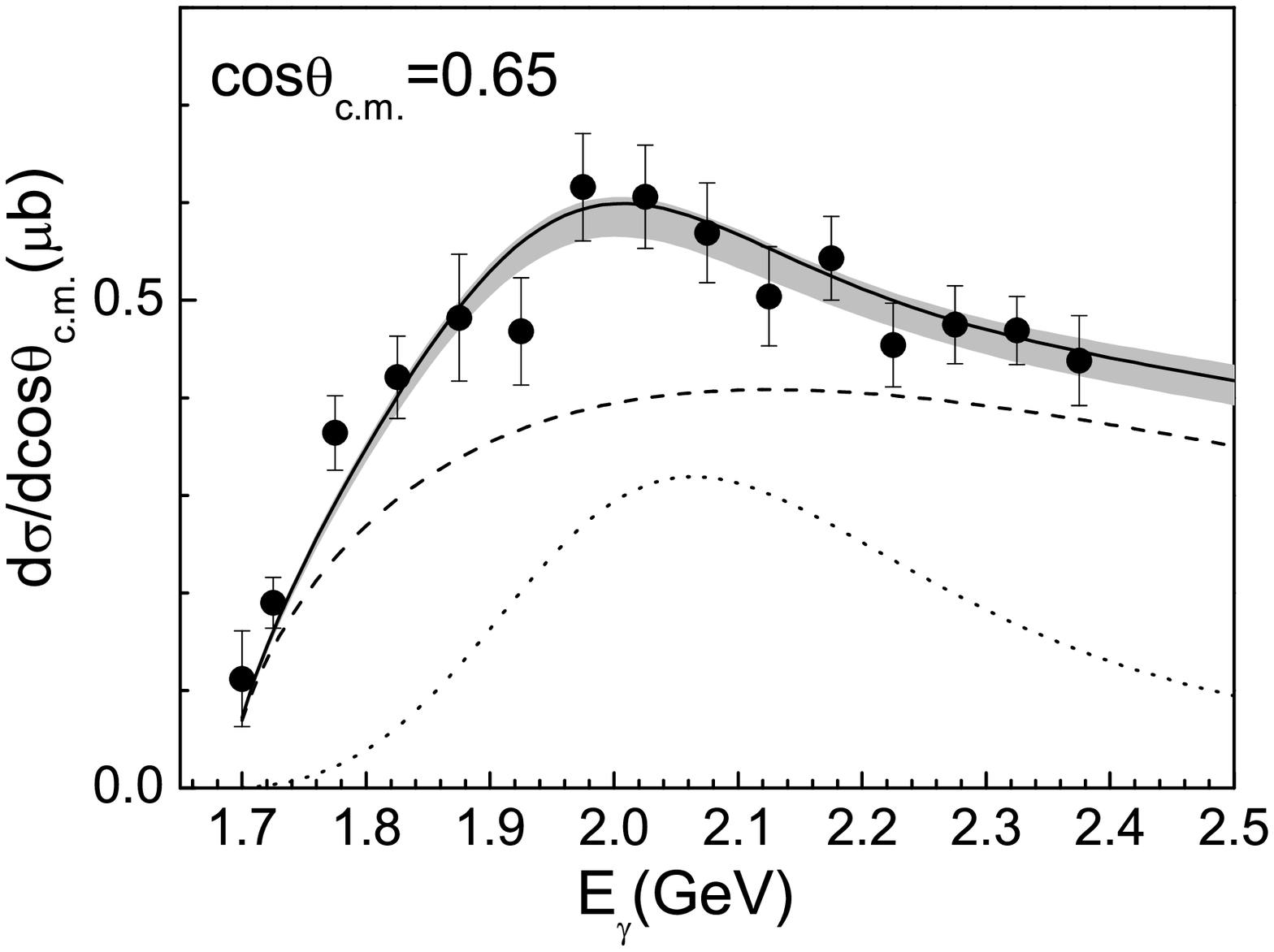}
\includegraphics[scale=0.23]{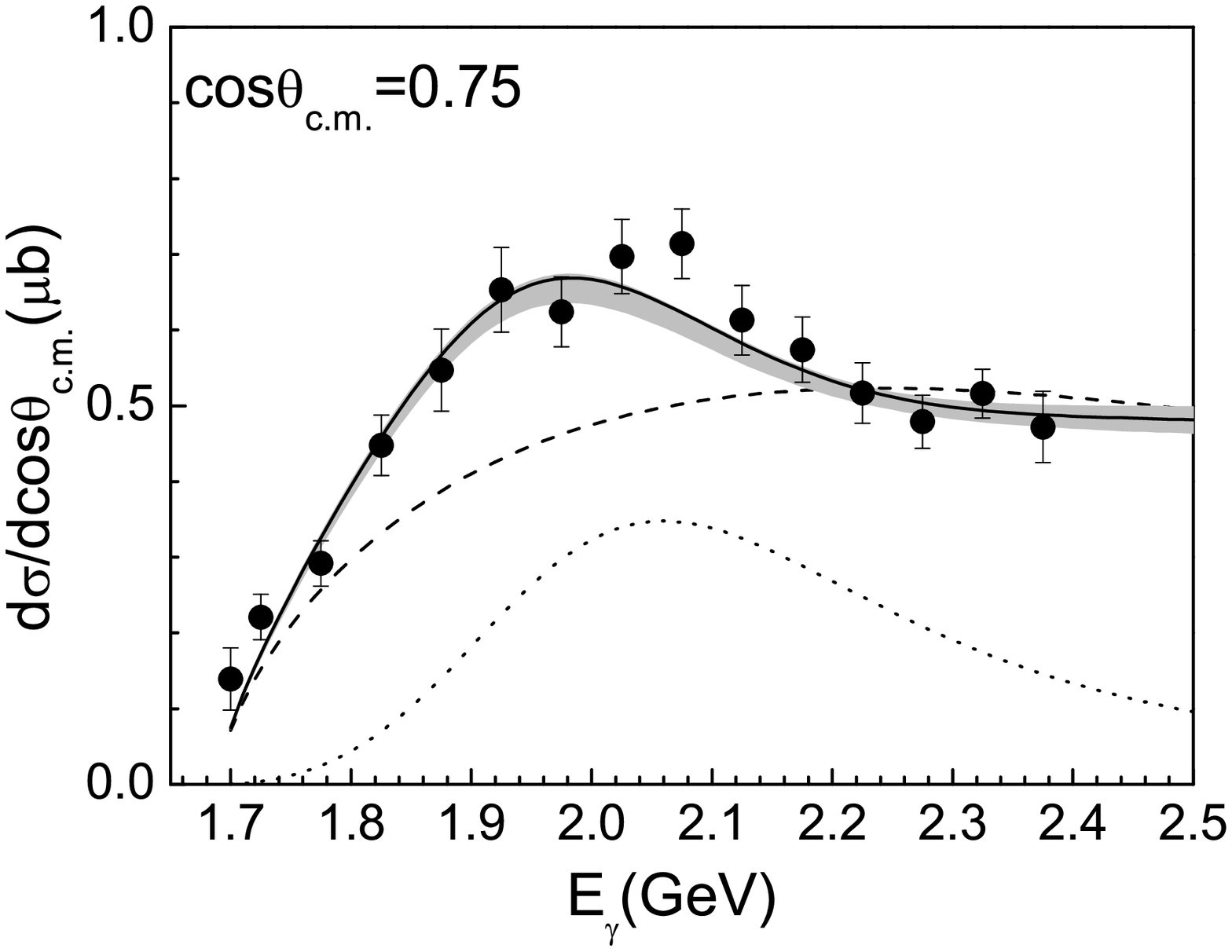}
\includegraphics[scale=0.23]{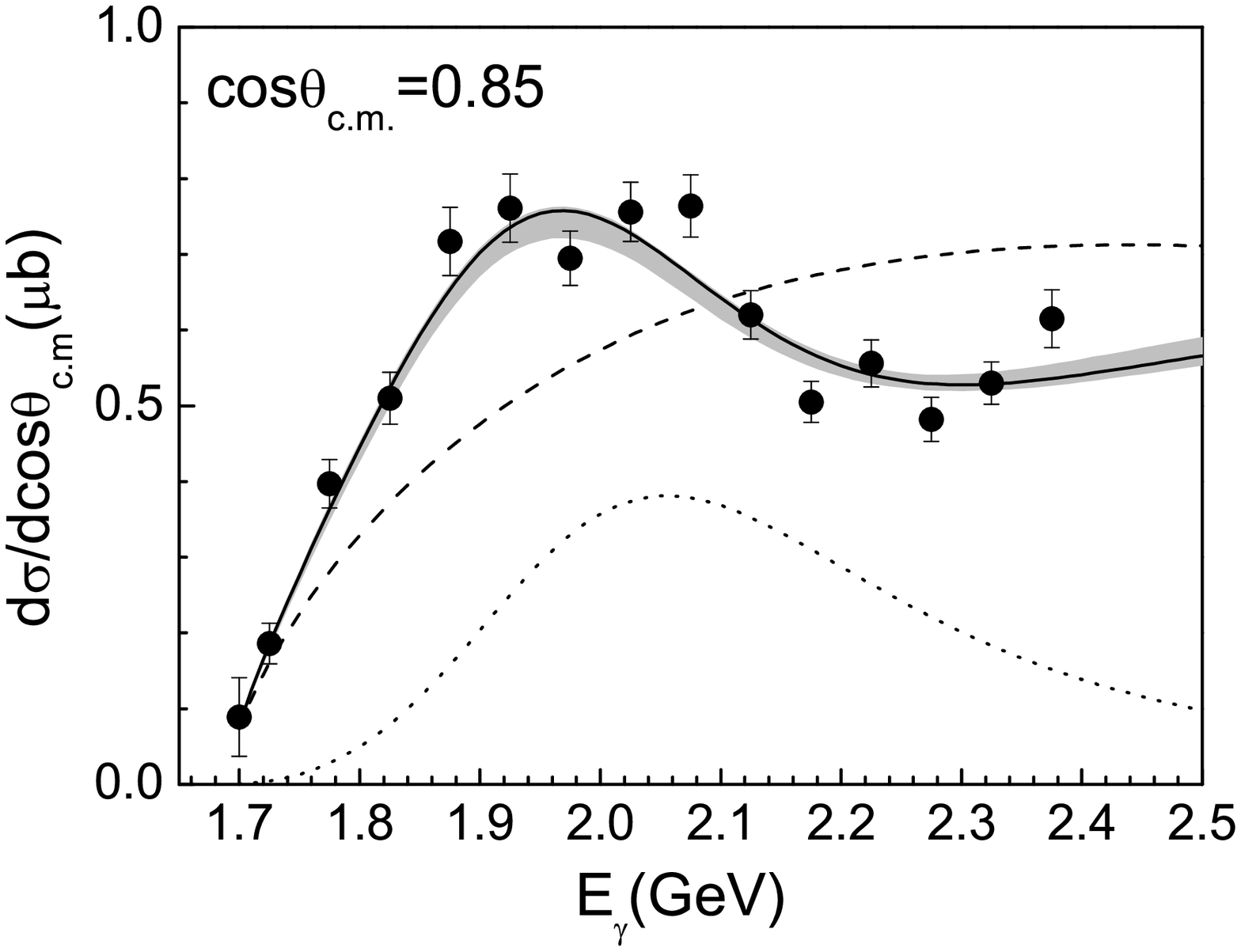}
\includegraphics[scale=0.23]{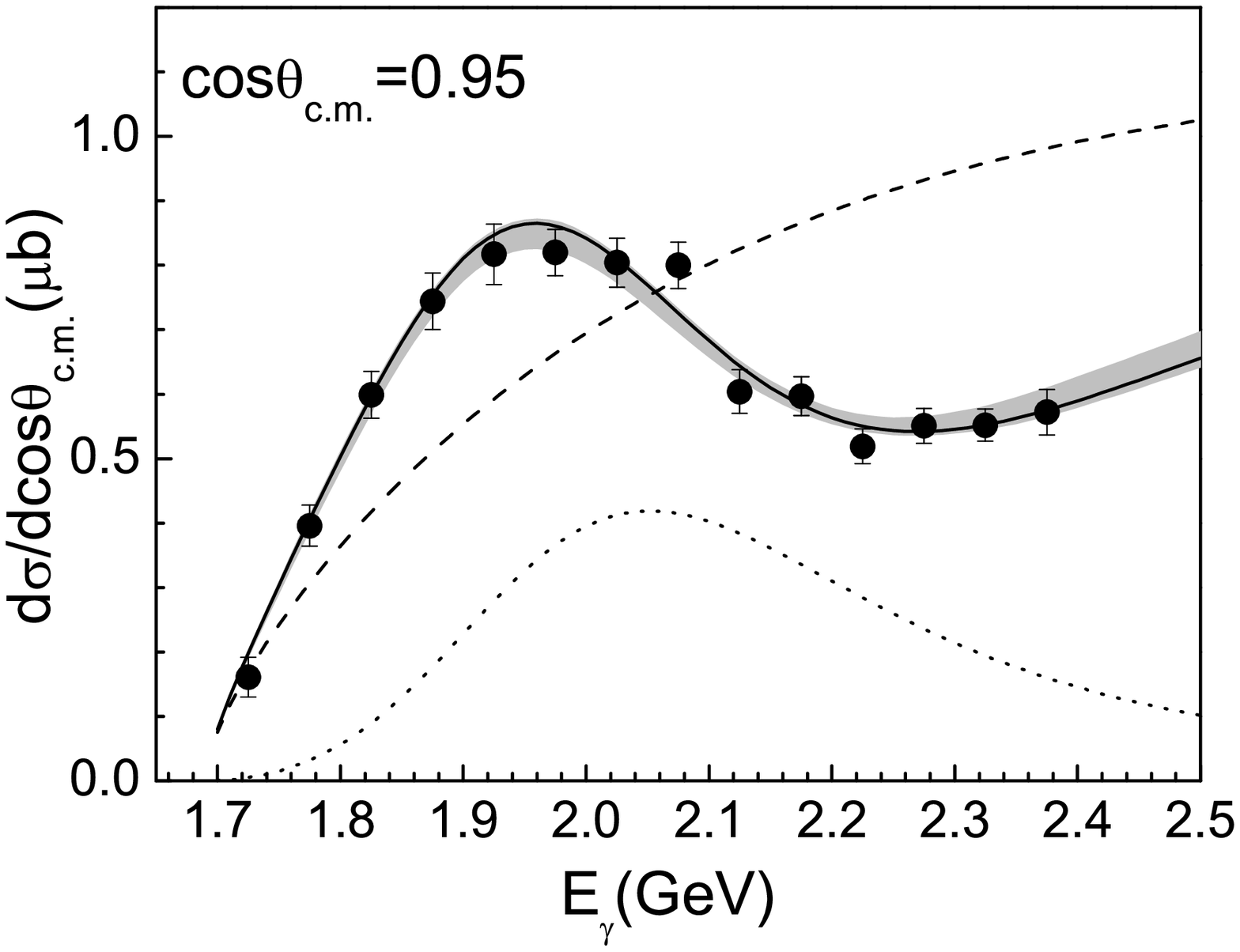}%
\caption{The $\gamma p \to K^+ \Lambda(1520)$ differential
$d\sigma/d(\cos\theta_{\rm c.m.})$ cross sections at forward $K^+$
angles compared with the LEPS data. Dashed and dotted lines show the
contributions from the background ($t-$channel $K^-$ exchange,
$s-$channel nucleon pole, and contact term) and $N^*(2120)$
resonance terms, respectively, while the solid line displays the
full result. For this latter curve we also show the statistical 68\%
confidence level (CL) band.}
\label{dcs-leps}%
\end{center}
\end{figure}

In Fig.~\ref{dcs-leps}, the theoretical calculations of the
differential cross section $d\sigma/d(\cos\theta_{\rm c.m.})$ as a
function of the photon beam energy $E_{\gamma}$ are shown. These
predictions are also compared to the experimental data taken from
LEPS collaboration~\cite{leps2}. The contributions from different
mechanisms of the model are separately shown. We see that the bump
structure in the differential cross section could be fairly well
described thanks to a significant contribution from the $N^*(2120)$.
It is worth to mention that in this calculation the contribution
from the $u-$channel $\Lambda(1115)$ pole has been neglected, since
its contribution is small at forward $K^+$ angels.

\begin{figure}
\begin{center}
\includegraphics[scale=0.75]{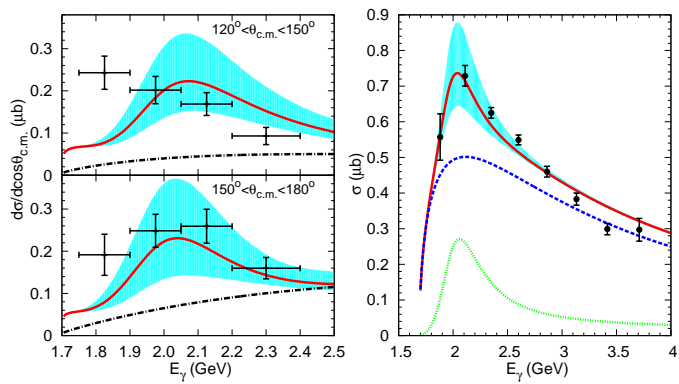}
\caption{The $\gamma p \to K^+ \Lambda (1520)$ differential cross
section (left panel) and total cross section (right panel) as a
function of the photon beam energy $E_{\gamma}$. Left panel:
black-dash-dotted curves show the $u-$channel $\Lambda (1115)$
contribution, while the red-solid lines show the full contributions.
Right panel: blue-dashed and green-dotted lines show the
contributions from the background and $N^*(2120)$ resonance terms,
respectively, while the red-solid line displays the results from the
full model. In both panels, we display the 68\% CL bands.}
\label{dcs-deg}%
\end{center}
\end{figure}

Next in Fig.~\ref{dcs-deg}, we show theoretical results for
differential cross sections at large $K^+$ angles and the total
cross section. Here, we also compare our predictions with the
experimental data from Refs.~\cite{leps1,Moriya:2013hwg}. In this
case, we pay attention not only to the contribution from $s-$channel
$N^*(2120)$ resonance but also to that from the $u-$channel
$\Lambda$ pole, since the $u-$channel contribution at the backward
angles could be important. As can be seen in the Fig.~\ref{dcs-deg},
the theoretical calculations provides a fair description of these
backward $K^+$ angular data thanks to the contribution from the
$\Lambda$ pole term in the $u-$channel. Furthermore, for the total
cross section, due to an important contribution from the
photo-excitation of the $N^*(2120)$ resonance and its subsequent
decay into a $\Lambda(1520)K^+$ pair, the theoretical results can
describe the CLAS data~\cite{Moriya:2013hwg} very well (see right
panel of Fig.~\ref{dcs-deg}). This mechanism is also important for
the bump structure in the LEPS differential cross section at forward
$K^+$ angles discussed in Fig.~\ref{dcs-leps}. Thus, one can
definitely take advantage of the important role played by this
resonant mechanism in the LEPS and CLAS data to better constrain
some of the $N^*(2120)$ properties, starting from its mere
existence.

\section{Study on $\pi^- p \to K^0 \Lambda(1520)$ and $pp \to pK^+ \Lambda(1520)$ reactions}

Since the $N^*(2120)$ resonance played an important role in the
$\gamma p \to K^+\Lambda(1520)$ reaction as discussed above, it may
also have important contributions to the $\pi^- p \to K^0
\Lambda(1520)$ and $pp \to pK^+ \Lambda(1520)$ reactions, which has
also been studied in Ref.~\cite{xieliu}.

For the $\pi^- p \to K^0 \Lambda(1520)$ reaction, the differential
cross section in the $\rm c.m.$ frame can be expressed as

\begin{equation}
{d\sigma \over d{\rm cos}\theta}={1\over 32\pi s}{
|\vec{p_3}^{\text{c.m.}}| \over |\vec{p_1}^{\text{c.m.}}|} \left (
{1\over 2}\sum_{s_{\Lambda^*},s_p}|T|^2 \right ). \nonumber
\end{equation}
In the equation above $\theta$ denotes the angle of the outgoing
$K^0$ relative to beam direction in the $\rm c.m.$ frame, while
$\vec{p_1}^{\text{c.m.}}$ and $\vec{p_3}^{\text{c.m.}}$ are the
three-momenta of the initial $\pi^-$ and final $K^0$ mesons,
respectively. The total invariant scattering amplitude $T$ is given
by,
\begin{equation}
T=T_s + T_t + T_u  + T_R \, ,
\end{equation}
where the contributions from the $s-$channel nucleon pole,
$t-$channel $K^*$ exchange, $u-$channel $\Sigma^+$ and $s-$channel
$N^*(2120)$ terms are considered.

\begin{figure}[htbp]
\begin{center}
\includegraphics[scale=0.45]{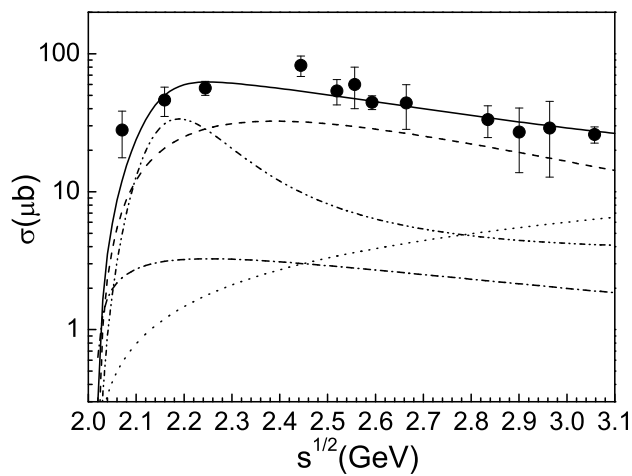}
\caption{Total cross sections vs the invariant mass
$s^{\frac{1}{2}}$ for the $\pi^- p \to K^0 \Lambda(1520)$ reaction.
The curves stand for the contributions from the $s-$channel nucleon
pole (dashed), $t-$channel $K^*$ (dotted), $u-$channel $\Sigma^+$
term (dash-dotted) and the $s-$channel $N^*(2080)$ terms
(dash-dot-dotted) and the total contribution (solid).}
\label{piptcs}
\end{center}
\end{figure}

With the parameters for the $N^*(2120)K\Lambda(1520)$ strong
couplings that were obtained from the $\gamma p \to K^+
\Lambda(1520)$ reaction, the role of the $N^*(2120)$ resonance in
the $\pi^- p \to K^0\Lambda(1520)$ reaction has been investigated.
Theoretical results for the total $\pi^- p \to K^0 \Lambda(1520)$
cross section are shown in Fig.~\ref{piptcs}, and compared with the
data taken from Ref.~\cite{pipdata}. The solid lines represent the
full results, while the contributions from the $s-$channel nucleon
pole, $t-$channel $K^*$ exchange, $u-$channel $\Sigma^+$ and
$s-$channel $N^*(2120)$ terms are shown by the dashed, dotted,
dot-dashed, and dash-dot-dotted lines, respectively. We see that we
can describe the experimental data of total cross sections quite
well, while the $s-$channel nucleon pole and $N^*(2120)$ resonance
and also the $u-$channel $\Sigma^+$ exchange give the dominant
contributions below $\sqrt{s}=2.4$ GeV. The $t-$channel $K^*$
exchange diagram gives a minor contribution.

\begin{figure}[htbp]
\begin{center}
\includegraphics[scale=0.5]{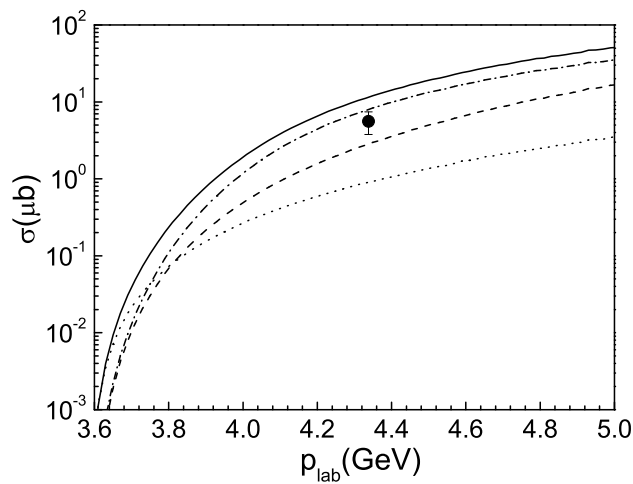}
\caption{Total cross section vs proton beam momentum ${\rm p_{lab}}$
for the $pp \to pK^+\Lambda(1520)$ reaction. The dashed, dotted, and
dash-dotted lines stand for contributions from nucleon pole,
$\Sigma^+$ pole and $N^*(2080)$ resonance, respectively. The total
contribution is shown by the solid line.} \label{pptcs}
\end{center}
\end{figure}

For the $pp \to pK^+ \Lambda(1520)$ reaction, the total cross
section versus the beam momentum (${\rm p_{lab}}$) of the proton is
calculated by using a Monte Carlo multi-particle phase space
integration program. The results for beam momentum ${\rm p_{lab}}$
from just above the production threshold $3.59$ GeV to $5.0$ GeV are
shown in Fig.~\ref{pptcs}. The dashed, dotted, and dash-dotted lines
stand for contributions from nucleon pole, $\Sigma^+$ pole and
$N^*(2120)$ resonance, respectively. The total contribution is shown
by the solid line. From Fig.~\ref{pptcs}, we mechanisms see that the
contribution from the $u-$channel $\Sigma^+$ exchange is dominant
very close to threshold, but, when the beam energy increases, the
contributions from the $s-$channel nucleon pole and the $N^*(2120)$
resonance turn to be very important.

It is worth to note that our predictions for the total $pp \to
pK^+\Lambda(1520)$ cross section, at ${\rm p_{lab}} = 3.65$ GeV, is
$0.01 \mu b$, which is $20$ times smaller than the experimental
upper limit of $0.2 \mu b$ as measured by the COSY-ANKE
Collaboration~\cite{zychorcosy}. This shows that our model
predictions are consistent with the experimental results. Moreover,
the total cross section of the $pp \to pK^+\Lambda(1520)$ reaction
has been also measured with HADES~\cite{l1520data} at GSI for a
kinetic energy ${\rm T_p} = 3.5$ GeV (corresponding to ${\rm
p_{lab}} = 4.34$ GeV). The result is $5.6\pm1.1\pm0.4^{+1.1}_{-1.6}
~\mu$b, as shown in Fig.~\ref{pptcs}, this is to be compared with
our theoretical result, $11.5 ~\mu$b. However, if we modify the cut
off parameters $\Lambda_{\pi}$ and $\Lambda^*_{\pi}$ from 1.3 GeV to
1.0 GeV, we get $\sigma = 5.45 ~\mu$b, which would be in agreement
with the experimental data. However, it does not make sense to fit
only one data point. So we still keep $\Lambda_{\pi} =
\Lambda^*_{\pi}$ = 1.3 GeV as used in many previous
works~\cite{xiezou}. We should also mention that, in the present
calculation, we did not include the $\Lambda(1520) p$
final-state-interaction (FSI), which can increase the results even
by a factor of 10 at the very near threshold region, similarly to
the important role played by $\Lambda p$ FSI in the $pp \to
pK^+\Lambda$ reaction~\cite{Xie:2011me}. We ignore this effect
because there are no experimental data on this reaction and also
very scarce information about the $\Lambda(1520) p$ FSI.

\begin{figure}[htbp]
\begin{center}
\includegraphics[scale=0.7]{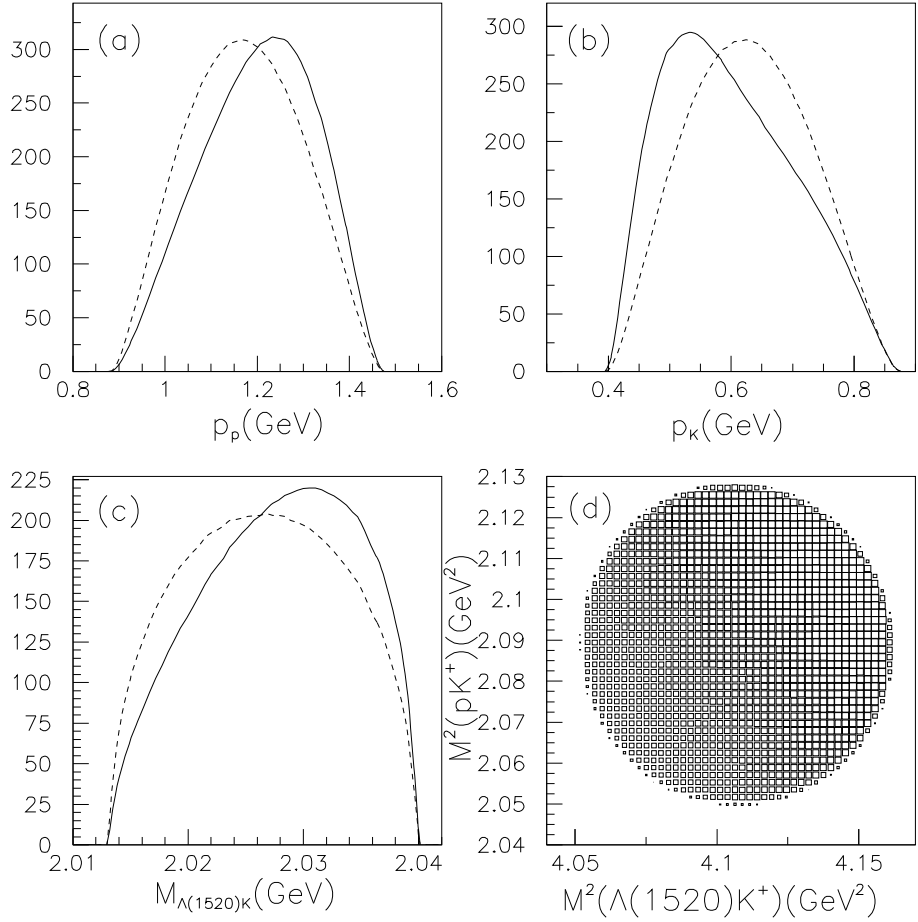}
\caption{Momentum distribution (arbitrary units), invariant mass
spectrum (arbitrary units), and Dalitz Plot for the $pp \to pK^+
\Lambda(1520)$ reaction at beam momentum ${\rm p_{lab}} = 3.67$ GeV
compared with the phase space distribution. The dashed lines are
phase space distributions, while the solid lines display full
results from our model.} \label{plab367}
\end{center}
\end{figure}

Furthermore, the corresponding momentum distributions of the final
proton and $K^+$ meson, the $K\Lambda(1520)$ invariant mass
spectrum, and also the Dalitz Plot for the $pp \to pK^+
\Lambda(1520)$ reaction at beam momentum ${\rm p_{lab}} = 3.67$ GeV,
which is accessible for DISTO Collaboration~\cite{baledisto}, are
calculated and shown in Fig.~\ref{plab367}(a),
Fig.~\ref{plab367}(b), Fig.~\ref{plab367}(c), and
Fig.~\ref{plab367}(d), respectively. The dashed lines are just phase
space distributions, while, the solid lines are full results from
our model. From Fig.~\ref{plab367}, we see that even at ${\rm
p_{lab}} = 3.67$ GeV, there is a clear bump in the $K\Lambda(1520)$
invariant mass distribution, which is produced by the contribution
of the $N^*(2120)$ resonance.

\section{Summary}

We have reviewed the role of $N^*(2120)$ resonance in the associate
$K\Lambda(1520)$ photon and hadronic productions at low energies
within an effective Lagrangian approach and the isobar model. In
addition to the contact, $t-$channel $\bar K$ exchange, and
$s-$channel nucleon pole contributions, the contributions from the
$u-$channel $\Lambda(1115)$ hyperon pole term and $N^* (2120)$
resonance are also considered. The results show that when the
contributions from the $N^*(2120)$ resonance and the $\Lambda(1115)$
are taken into account, both the new CLAS~\cite{Moriya:2013hwg} and
the previous LEPS data~\cite{leps1,leps2} for the $\gamma p \to
K^+\Lambda(1520)$ reaction can be simultaneously described.

Actually, we find an overall good description of the data, both at
forward and backward $K^+$ angles, and for the whole range of
measured $\gamma p$ invariant masses. The contribution of the
$u-$channel $\Lambda (1115)$ pole term produces an enhancement at
backward angles, and it becomes more and more relevant as the photon
energy increases, becoming essential above $W \ge 2.35$ GeV and
$\cos\theta_{\rm c.m.} \le -0.5$. On the other hand, the CLAS data,
clearly support the existence of a spin-parity $J^P = 3/2^-$ nucleon
resonance with a mass around 2.1 GeV, a width of at least 200 MeV
and a large partial decay width into $\Lambda(1520)K$. These
characteristics could be easily accommodated within the constituent
quark model results of Simon Capstick and W. Roberts of
Ref.~\cite{simonprd58}. Such resonance might be identified with the
two stars PDG $N^*(2120)$ state, which would confirm previous
claims~\cite{xiejuan,hejun} from the analysis of the bump structure
in the LEPS differential cross section at forward $K^+$ angles
discussed in Fig.~\ref{dcs-leps}.

On the other hand, motivated by the study of the $\gamma p \to
K^+\Lambda(1520)$ reaction, the role of $N^*(2120)$ has also been
investigated in the $\pi^- p \to K^0 \Lambda(1520)$ and $pp \to
pK^+\Lambda(1520)$ reactions. The results show that the contribution
from the $u-$channel $\Sigma^+$ exchange is dominant in the very
near threshold region, but, when the beam energy increases, the
contributions from $s-$channel nucleon pole and $N^*(2120)$
resonance turn to be very important. Furthermore, the invariant mass
distribution and the Dalitz Plot are also predicted which can be
tested by the future experiments.

\section*{Acknowledgments}

This work is partly supported by DGI and FEDER funds, under
contracts FIS2011-28853-C02-01 and FIS2011-28853-C02-02 the Spanish
Ingenio-Consolider 2010 Program CPAN (CSD2007-00042), Generalitat
Valenciana under contract PROMETEO/2009/0090 and by the National
Natural Science Foundation of China under grants 11105126 and
10905046. We acknowledge the support of the European
Community-Research Infrastructure Integrating Activity "Study of
Strongly Interacting Matter" (acronym HadronPhysics2, Grant
Agreement n. 227431) under the Seventh Framework Programme of EU.
Work supported in part by DFG (SFB/TR 16, "Subnuclear Structure of
Matter").


\end{document}